\documentclass[prl,nofootinbib,twocolumn]{revtex4}

\usepackage{color}
\usepackage{graphicx}
\usepackage{amsmath}

\begin{document}

\title{Measuring the speed of cosmological gravitational waves}

\author{Marco Raveri$^{1,2}$, Carlo Baccigalupi$^{1,2}$, Alessandra Silvestri$^{1,2,3}$ and Shuang-Yong Zhou$^{1,2}$}

\smallskip
\affiliation{$^{1}$ SISSA - International School for Advanced Studies, Via Bonomea 265, 34136, Trieste, Italy \\
\smallskip
$^{2}$ INFN, Sezione di Trieste, Via Valerio 2, I-34127 Trieste, Italy \\
\smallskip
$^{3}$ INAF-Osservatorio Astronomico di Trieste, Via G.B. Tiepolo 11, I-34131 Trieste, Italy}

\begin{abstract}
In general relativity gravitational waves propagate at the speed of light, however in alternative theories of gravity that might not be the case.
We study the effects of a modified speed of gravity, $c_T^2$,  on the B-modes of the Cosmic
Microwave Background (CMB) anisotropy in polarisation.
We find that a departure from the light speed value would leave a
characteristic imprint on the BB spectrum part induced by tensors,
manifesting as a shift in the angular scale of its peaks.
We derive constraints by using the available {\it Planck} and BICEP2
datasets showing how $c_T^2$ can be measured, albeit obtaining
weak constraints due to the overall poor accuracy of the current BB
power spectrum measurements.
The present constraint corresponds to $c_T^2 = 1.30 \pm 0.79$  and
$c_T^2< 2.85$ at $95\%$ C.L. by assuming a power law primordial tensor
power spectrum and $c_T^2<2.33$ at $95\%$ C.L. if the running of the
spectral index is allowed.
We derive forecasts for the next generation CMB satellites, which we
find capable of tightly constraining $c_T^2$ at percent level, comparable with bounds from binary pulsar measurements, largely due to the absence of degeneracy with other
cosmological parameters.
\end{abstract}
\maketitle
Despite decades of intensive effort, gravitational waves (GWs) have yet to be observed directly.  While the situation may change in the coming decade~\cite{Marka:2010zza}, a new window  has been opened by CMB experiments that have recently detected the B-modes of polarisation~\cite{Ade:2014xna,Ade:2014afa}, offering an indirect measurement of cosmological GWs (tensor modes). While on small angular scales the BB-power spectrum is dominated by the lensing of the CMB, on larger scales, the B-modes of polarisation are primarily produced by tensor modes and give an insight onto primordial GWs~\cite{Zaldarriaga:1996xe}. \\
\indent  In general relativity, short-wavelength GWs follow the null geodesics of the background, thus their propagation speed equals the speed of light on a flat background. However in alternative theories addressing the phenomenon of cosmic acceleration, in Ho$\check{\rm r}$ava-Lifshitz~\cite{Horava:2009uw,Bogdanos:2009uj} gravity and, more generally, in Lorentz-violating theories~\cite{Rubakov:2004eb, Dubovsky:2004sg,Liberati:2013xla}, the speed of gravity may deviate from that of light. For instance, some of the generalized scalar-tensor models within the Horndeski family~\cite{Horndeski:1974wa,Deffayet:2011gz}, like the covariant galileon involving certain derivative couplings, are expected to modify the tensor propagation speed~\cite{Amendola:1993uh,Kobayashi:2011nu,Gao:2011qe,DeFelice:2011uc};  quantum gravity effects may modify the dispersion relation  of GWs~\cite{Cai:2014hja}; or the graviton may have a mass which prevents it from behaving light-like~\cite{Dubovsky:2009xk}. Massive gravity has the 
added complication that diffeomorphism invariance is explicitly broken, therefore in this paper we will focus on variations of $c_T$ and will not consider a mass term~\cite{Dubovsky:2009xk,Gumrukcuoglu:2012wt}. \\
\indent A direct measurement of the speed of GWs could be achieved comparing the arrival times of light and gravitational wave signals from a distant astronomical source~\cite{Will:2014kxa}. This has not been possible yet, however indirect, local observations of gravitational radiation seem to suggest that its propagation speed, at the current epoch, is close to the speed of light. For instance,  accurate measurements of binary pulsar timing indicate that the sound speed of GWs should not deviate from the general relativistic value by more than 1\%~\cite{Will:2014kxa}; the latter bound assumes that energy is lost via GWs; models of modified gravity  might however imply also a loss of energy via scalar radiation associated to an additional d.o.f., possibly modifying this bound~\cite{Silvestri:2011ch,deRham:2012fg,Brax:2013uh}. \\
\indent In this paper we focus on the B-modes of CMB polarisation and show how they offer a novel, {\it independent} way to measure the propagation speed of GWs at the time of recombination.  As we will show,  variations of $c_T$ affect the BB power spectrum in a unique way which makes it orthogonal with other cosmological parameters. Interestingly, as we will discuss, the B-modes are only sensitive to the modifications of the dispersion relation of the graviton around the time of recombination, therefore they are complementary to bounds from binary pulsars, allowing a combined constraint on the time variation of $c_T$. We derive bounds from current data, which we find to have limited constraining power, as well as forecasts from upcoming and future, cosmic variance limited, CMB experiments.\\
\begin{figure*}[t]
\includegraphics[width=0.90\textwidth]{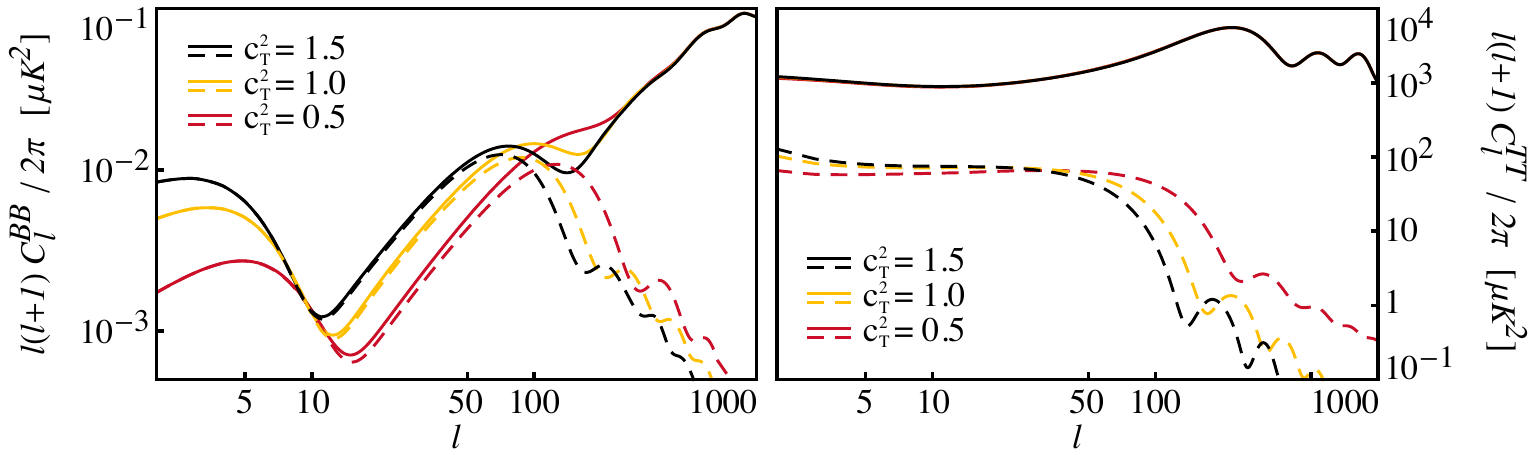}
\caption{{\it Left:} the total B-mode polarisation power spectrum (solid lines) and its component due to tensor perturbations (dashed lines). {\it Right:} the total CMB temperature power spectrum (solid lines) and its tensor component (dashed lines). In both panels different colors correspond to different values of the speed of GWs. The other cosmological parameters are fixed to the best fit of the {\it Planck} + BICEP2 datasets.}
\label{fig:PowerSpectra}
\end{figure*}
\indent On a flat Friedmann-Robertson-Walker background one can use the rotational and translational symmetries  to decompose the metric perturbations into scalar, vector and tensor components. 
We are only interested in the tensorial part:
\begin{equation}
{\rm d}s^2 = - a(\tau)^2[ {\rm d} \tau^2 + (\delta_{ij}+h_{ij}) {\rm d}x^i {\rm d}x^j ]   \,,
\end{equation}
where $h_{ij}$ satisfy $\partial_i h_{ij} =0$ and $h_{ii}=0$. We shall consider a linear perturbation theory that also satisfies the gauge symmetry $h_{ij}\to h_{ij}+\partial_{(i}\epsilon_{j)}$, where $\epsilon_i$ is a generic function of the coordinates. \\
\indent We can write down the most general quadratic action for $h_{ij}$ that is ghost free and satisfies the symmetries mentioned above: 
\begin{align} \label{Eq:TensorAction}
S_T^{(2)} = \frac{1}{8} \int {\rm d}\tau\, {\rm d}^3 x \,  a^2 M_P^2(\tau) \left({h}'_{ij}{h}'_{ij} - c^2_T(\tau) \partial_k h_{ij} \partial_k h_{ij} \right)  \,,
\end{align}
where matter sources are assumed to be minimally coupled to the metric. 
Generally higher order gradient terms $h_{ij}\partial^{2n} h_{ij}(n>1)$ should also be included in the action~(\ref{Eq:TensorAction}) but we have neglected them as we are interested in the low energy phenomenology~\cite{Gumrukcuoglu:2012wt}.
The function $M_P^2(\tau)$ plays the role of the Planck mass which is allowed by the above mentioned symmetries to vary in time.
We fix its value to $M_P^2\equiv(8\pi G)^{-1}$ since time variations of this quantity are constrained by a large number of complementary observations~\cite{Will:2014kxa,Will:2005va,Clifton:2011jh} while polarisation observables are expected to depend weakly on this quantity.
\begin{figure*}[t]
\includegraphics[width=0.90\textwidth]{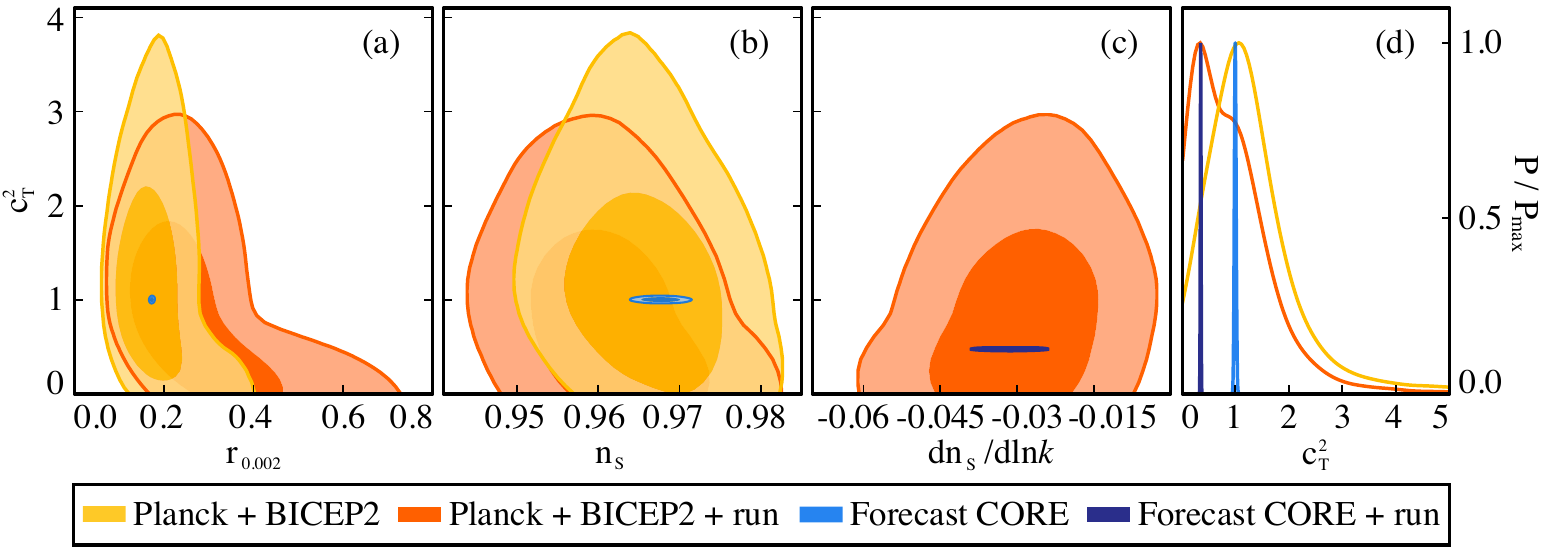}
\caption{{\it Left:} The marginalized joint likelihood for the GWs speed of sound $c_T^2$  and parameters defining the primordial tensor power spectrum:  the tensor to scalar ratio $r_{0.002}$, the scalar perturbations spectral index $n_s$ and its running $dn_s/d{\rm ln}k$. Different colors correspond to different combinations of datasets and models as shown in the legend. The two different shades indicate the $68\%$ and the $95\%$ confidence regions. {\it Right:} Marginalized likelihoods of the tensor perturbations sound speed for the considered datasets and models.}
\label{fig:MarginalData}
\end{figure*}
We are then left with modifications of  the sound speed of tensor modes, so that the action~(\ref{Eq:TensorAction}) results in the following wave equation:
\begin{align}\label{Eq:TensorPropagation}
h''_{ij} + 2\frac{a'}{a} h'_{ij} - c_T^2 \partial_k\partial_k h_{ij}  = \frac{2}{M_P^2a^2} S_{ij} \,,
\end{align}
where $S_{ij}$ is the transverse and traceless component of the energy momentum tensor of the matter sector.
Varying $c_T^2$ changes the relevant dynamical scale of tensor perturbations from the effective cosmological horizon, corresponding to the case $c_T^2=1$ (in units of the speed of light), to the sound horizon.
For this reason the net effect on CMB spectra is an horizontal shift of the whole tensor induced component whose main peak moves at the angular scale of the GWs sound horizon at recombination as can be seen in figure~\ref{fig:PowerSpectra}. Notably the sources of the E and B-mode polarisation spectra are peaked at the recombination epoch~\cite{Zaldarriaga:1996xe} thus making them dependent on the dynamics of tensor perturbations at earlier times but limiting the impact of a later evolution. We have studied this effect numerically and found that a possible late time dependence of the GWs sound speed does not noticeably affect the polarisation observables which are in turn sensitive to its value around the recombination time. 
According to this result we have assumed $c_T^2$ to be constant throughout all cosmological epochs. Changing the speed of GWs in principle impacts also the reionization bump at large scales in the polarisation spectra. The effect is however less prominent than the shift of the recombination peak and its constraining power is also reduced by cosmic variance, which is stronger at those scales.
The effect of horizontal shifting of the tensor component of the CMB spectra can in principle be mimicked by a change in the cosmological expansion history which is however tightly constrained by the scalar part of the CMB itself~\cite{Ade:2013zuv,Antolini:2012kv} and many other observations.  Other cosmological parameters, especially those defining the primordial tensor power spectrum, are not expected to be degenerate 
with $c_T^2$. The tensor to scalar ratio, for example, shifts vertically the tensor part of CMB spectra while the spectral index primarily changes its shape.
More complicated models for the tensor primordial power spectrum could in principle be degenerate with $c_T^2$ but if inflationary consistency relations~\cite{Liddle:1992wi} are assumed then the scalar sector is expected to break this degeneracy. \\
\indent From figure~\ref{fig:PowerSpectra} we can also see that the effect of changing $c_T^2$ weakly influences the CMB temperature power spectrum because the tensor induced component is several orders of magnitude smaller than the scalar one for values of the tensor to scalar ratio not yet excluded by observations.
We have also investigated the influence of this effect on the E-mode polarisation spectrum and found it negligible.
The B-mode spectrum is instead greatly influenced by changes in the speed of GWs thus making this CMB observable the most suited for these studies. \\
\indent We use the recently released BICEP2~\cite{Ade:2014xna} data along with the {\it Planck} CMB temperature power spectrum measurements~\cite{Ade:2013kta} and  the WMAP low-$\ell$ polarisation spectra~\cite{Hinshaw:2012aka} to constrain the speed of sound of cosmological GWs. 
To forecast the precision at which this quantity will be measured by the next generation of CMB experiments we create simulated datasets adopting the specifications of the {\it Cosmic Origins Explorer} (CORE)~\cite{CoREWhitePaper} and the {\it Polarized Radiation Imaging and Spectroscopy Mission} (PRISM)~\cite{Andre:2013nfa} satellites. 
We perform a Markov Chain Monte Carlo analysis of both the current data and the simulated data using the publicly available CosmoMC package~\cite{Lewis:2002ah}; in the case of forecast this allows us to have a good handle on the degeneracies among cosmological parameters. We allow variation of the six baseline cosmological parameters of the $\Lambda$ Cold Dark Matter ($\Lambda$CDM) model, plus the running of the scalar spectral index and the amplitude of primordial cosmological GWs; we impose the  single field inflationary scenarios to relate the spectral index of tensors to the scalar one. \\
\begin{figure*}[t]
\includegraphics[width=0.90\textwidth]{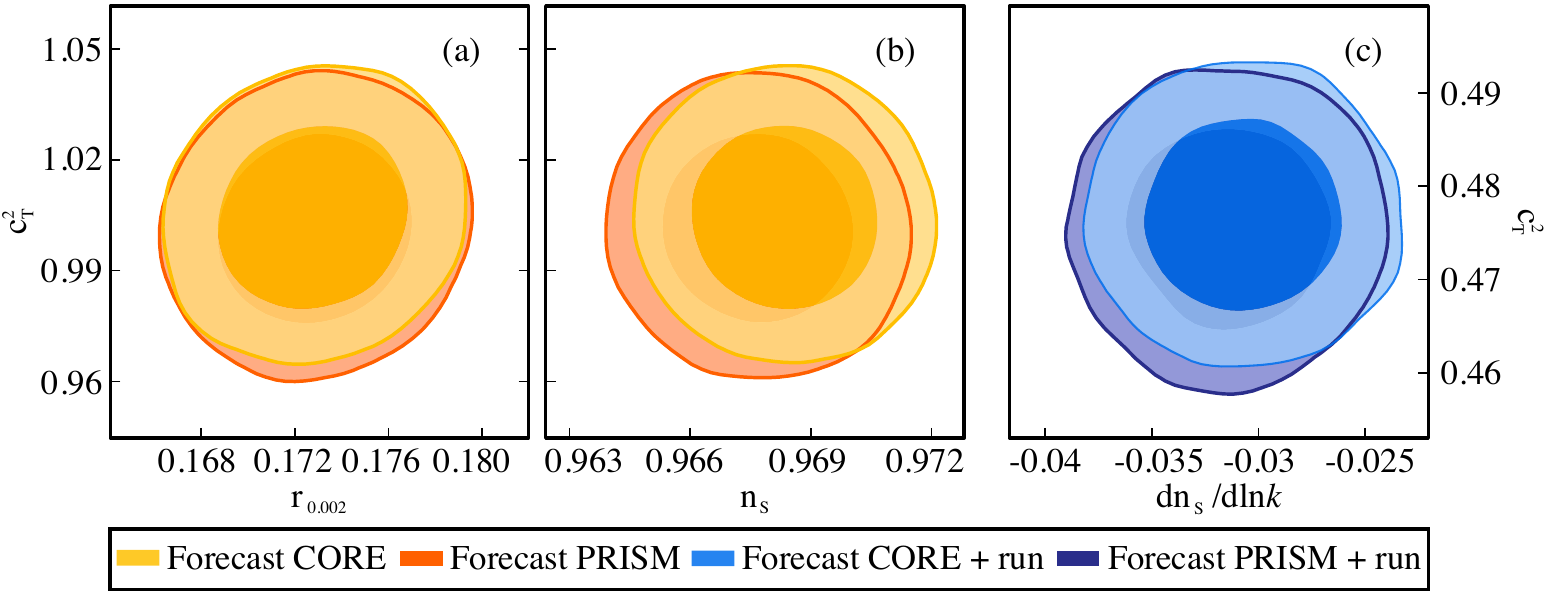}
\caption{The marginalized joint likelihood for the GWs speed of sound $c_T^2$,  the tensor to scalar ratio $r_{0.002}$, the scalar perturbations spectral index $n_s$ and its running $dn_s/d{\rm ln}k$. Different colors correspond to different instrumental specifications used in the forecast and different models as shown in the legend. The two different shades indicate the $68\%$ and the $95\%$ confidence regions.}
\label{fig:MarginalForecast}
\end{figure*}
\indent The results of this analysis for both current data and forecast are shown in figure~\ref{fig:MarginalData}, while in figure~\ref{fig:MarginalForecast}  we provide a zoom of the forecasts  of  the cosmological parameters most relevant for our analysis. From  panel (d)  of figure~\ref{fig:MarginalData} we can see that the marginalized likelihood of $c_T^2$ is peaked at its GR expected value, i.e. $c_T^2=1$ (in units of the speed of light) when considering a model without the running of the spectral index. 
Very high values of $c_T^2$ are excluded since they would move the tensor component of the B-mode spectrum to large scales, resulting in a poor fit of the measured data points. 
From panel (a) of figure~\ref{fig:MarginalData}, we can notice that there is a degeneracy  between $c_T^2$ and the tensor to scalar ratio assumed at a pivot scale of $0.002\, \text{Mpc}^{-1}$. 
The reason for this degeneracy is that those values of $c_T^2$ shift the GWs contribution to the spectrum toward smaller scales so that the only way to fit the data points is to change the spectrum amplitude. 
We can also see from panel (b) of figure~\ref{fig:MarginalData} that $c_T^2$ is weakly degenerate with the spectral index due to the poor constraining power of the BICEP2 measurements.  From the combination of the {\it Planck} and BICEP2 datasets we obtain the marginalized bound: $c_T^2 = 1.30 \pm 0.79$  and $c_T^2< 2.85$ at $95\%$ C.L..  \\
\indent If we allow a running of the primordial tensor power spectrum index the situation changes slightly. 
From the marginalized joint likelihood of $c_T^2$, $r_{0.002}$, $n_s$ and $\text{dn}_s/\text{dln}\,k$ in panels (a,b,c) of figure~\ref{fig:MarginalData}, we can see that $c_T^2$ is driven toward smaller values and this is further confirmed by its marginalized distribution in panel (d). 
The peak of the probability distribution of $c_T^2$ is found not to be at its GR value which is however not excluded. 
From the same figure we can see that as $c_T^2$ goes toward smaller values its degeneracy with $r_{0.002}$ is enhanced while it is not so pronounced with respect to the running of the spectral index, shown in panel (c), and $n_s$ itself, shown in panel (b).
Given the skewness of the marginal distribution of $c_T^2$ which is also cut at $c_T^2=0$ we report here only its upper bound: $c_T^2<2.33$ at $95\%$ C.L.. \\
\indent We now turn to the forecasts to investigate further these degeneracies and to evaluate our capability of constraining the speed of cosmological GWs with future generation surveys. Our results do not include any forecast on de-lensing capability, and thus represent rather conservative bounds in the adopted forecast setup. Indeed the CMB lensing signal represents the main contaminant for measurement of primordial GWs from the BB spectrum and the constraining power will improve accordingly to the capability of tracing this signal.
The results are shown in figure~\ref{fig:MarginalForecast}, with a fiducial model assumed to be the best fit one obtained with the {\it Planck} and BICEP2 datasets. 
We can clearly see that increasing the accuracy of B-mode polarisation observations removes all the degeneracy with the other cosmological parameters since the measurements would be able to disentangle the effect of horizontal shifting, due to changes in $c_T^2$, from the vertical shifting induced by varying $r_{0.002}$ or the shape changes due to $n_s$ and the running of the spectral index. 
As a result we can say that the parameter $c_T^2$, quantifying the speed of GWs at recombination, is orthogonal to other cosmological parameters, as it is theoretically expected. We can also notice that considering an instrument with higher precision like PRISM, does not improve significantly on the determination of $c_T^2$ with respect to CORE, since the effect is seen at degree angular scales where both of the considered experiments are cosmic variance limited. \\
\indent Overall,  we see that the next generation of CMB experiments will constrain the speed of cosmological GWs with a $1\%$ accuracy independently of the assumed shape of the primordial power spectrum. \\
\indent To summarize, we have considered a varying sound speed for cosmological GWs and its effects on the  power spectrum of B-modes of the Cosmic Microwave Background (CMB) anisotropies polarisation. We have derived the constraints on this quantity using the present CMB data from {\it Planck} and the BB power spectrum on the degree angular scale as reported recently by the BICEP2 experiment. Moreover, we have derived the projected 
constraints which will be within reach of the future generation, polarisation dedicated CMB satellites. \\
\indent We have found that a departure of $c_T$ from the speed of light has an apparent projection effect on the characteristic peak of the BB power spectrum at the degree scale, which corresponds to the part produced by cosmological GWs. We have identified the origin of this feature in the effective ``re-scaling'' of the gradient term of the wave equation~(\ref{Eq:TensorPropagation}), affecting the horizon re-entry time for the tensor component, and thus the location of the BB peak in the angular domain. \\
\indent Despite of the high constraining power of {\it Planck} on $\Lambda$CDM parameters, the claimed statistical significance in current measurements of B-modes results in a broad constraint on the GWs sound speed, $c_T$, corresponding to $c_T^2 = 1.30 \pm 0.79$  and $c_T^2< 2.85$ at $95\%$ C.L. 
by assuming a power law primordial tensor power spectrum; $c_T^2<2.33$ at $95\%$ C.L. is obtained if a running of the spectral index is allowed.  Since the effect of $c_T^2$ is rather orthogonal with that of other cosmological parameters, we have established the ultimate constraining power by adopting the specifications of the future proposed CMB satellites. We have found that those are indeed capable 
of resolving the parameter space into a neat constraint on $c_T$, without degeneracies with other parameters, down to a percent level. We observe that such a constraining power is almost competitive with the one from observations of binary pulsar timing. \\
\indent These results confirm the relevance of CMB polarisation measurements in exploring fundamental physics. Current sub-orbital probes are expected to improve substantially on the present constraints on the  propagation velocity of GWs. As we have shown, this quantity is independent from other $\Lambda$CDM parameters, and consequently the ultimate precision,  within reach of future, polarisation dedicated CMB satellites, will be at the level of the best probes which have been conceived so far. \\
\indent We conclude by outlining the expected progress in the near future. 
The {\it Planck} collaboration will publish results including polarisation within the end of the year. At the same time, the progress in the observations at degree and arcminute (e.g. BICEP2~\cite{Ade:2014xna} and PolarBear~\cite{Ade:2014afa}) scales, are expected to contribute substantially to the measurement of the tensor cosmological component constraining the effects which contribute in shaping the BB spectrum of CMB polarisation anisotropies.
\vskip 10pt
We acknowledge helpful discussions with Bin Hu, Stefano Liberati, Levon Pogosian and Daniele Vernieri.
We are grateful to Matteo Martinelli for useful conversations and help with CMB forecasts. 
We are particularly in debt with Noemi Frusciante for collaboration in the early stages of the work.  
AS acknowledges support from a SISSA Excellence Grant and partial support from the Italian Space Agency through the ASI contracts Euclid-IC (I/031/10/0). MR and AS acknowledge partial support from the INFN-INDARK initiative. SYZ acknowledges partial financial support from the European Research Council under the European Union’s Seventh Framework Programme (FP7/2007-2013) / ERC Grant Agreement n. 306425 “Challenging General Relativity”.
\vskip 10pt
\noindent {\it Note added.---}While this paper was in preparation, a related paper was posted on arxiv.org~\cite{Amendola:2014wma} commenting on similar ideas. Our work focuses on the detectability of $c_T$ while providing forecasts for next generation CMB experiments.

\end{document}